\documentclass[aps,prl,preprint,superscriptaddress,showpacs]{revtex4}
\usepackage{graphicx}
\usepackage{bm}
\begin{document}

\title{Layering and position-dependent diffusive dynamics of confined fluids}

\author{Jeetain Mittal}
\email[]{jeetain@helix.nih.gov}
\affiliation{Laboratory of Chemical Physics, National Institute of Diabetes and Digestive and Kidney Diseases, National Institutes
of Health, Bethesda, Maryland 20892-0520, USA}

\author{Thomas M. Truskett}
\email[]{truskett@che.utexas.edu}
\affiliation{Department of Chemical Engineering and Institute for Theoretical Chemistry,
The University of Texas at Austin, Austin, Texas 78712-0231, USA}

\author{Jeffrey R. Errington}
\email[]{jerring@buffalo.edu}
\affiliation{Department of Chemical and Biological Engineering, University
at Buffalo, The State University of New York, Buffalo, New York 14260-0231, USA}

\author{Gerhard Hummer}
\email[]{hummer@helix.nih.gov}
\affiliation{Laboratory of Chemical Physics, National Institute of Diabetes and Digestive and Kidney Diseases, National Institutes
of Health, Bethesda, Maryland 20892-0520, USA}

 \date{\today}

\begin{abstract}
We study the diffusive dynamics of a hard-sphere fluid confined between parallel smooth hard walls.  The position-dependent diffusion coefficient normal to the walls is larger in regions of high local packing density.  High density regions also have the largest available volume, consistent with the fast local diffusivity. Indeed, local and global diffusivities as a function of the Widom insertion probability approximately collapse onto a master curve. Parallel and average normal diffusivities are strongly coupled at high densities and deviate from bulk fluid behavior.
\end{abstract}

\maketitle

\section{}
Beginning with Einstein's theory of Brownian motion, it has been realized that diffusion provides an excellent approximation to the motion of molecules in a bulk liquid for times much longer than intervals between molecular collisions.  In molecularly confined systems, the situation is more complicated. On one hand, particle structures emerge that result in a spatially inhomogeneous density profile; and on the other hand, the relaxation time of local density fluctuations may become faster than the time required for particle motions to become ``diffusive."  Alternatively, long-lived correlations can prevent the system from entering into a diffusive regime altogether, as in single-file transport~\cite{Percus2003}.  Nevertheless, in simulations of partially confined systems such as  fluids in two-dimensional (2d) slit pores, diffusion is indeed observed parallel to the (quasi) infinite confining planes~\cite{Ted1985,Mittal2006}.  However, the situation is less clear for motions in the perpendicular direction.  Even if diffusion were a useful description of the single-particle motions, one would expect the diffusion coefficient to be spatially inhomogeneous.  That, combined with the spatially varying density profile and the confining boundaries, essentially eliminates the usual way of estimating diffusion coefficients from the mean square displacement as a function of time~\cite{Liu2004}.  As a consequence, diffusion in highly-confined environments has largely remained unexplored, despite its relevance for micro- and nanofluidic devices~\cite{Bocquet2007}.

We use a recently proposed propagator-based formalism to estimate the position-dependent diffusion coefficients self-consistently from simulation trajectory data~\cite{Gerhard2005}. For diffusion, the propagator (or Green's function) $G(z,\Delta t|z^{\prime},0)$ for single-particle displacements along the coordinate $z$ normal to the confining walls is assumed to satisfy the Smoluchowski diffusion equation,
 \begin{equation}
\frac{\partial G}{\partial t} = \frac{\partial}{\partial z}\left\lbrace D_\perp(z)\text{e}^{-\beta F(z)}\frac{\partial}{\partial z}[\text{e}^{\beta F(z)}G]\right\rbrace
\label {smoluchowski}
\end{equation}
with $\beta = 1/k_{\text B} T$, $k_{\text B}$ Boltzmann's constant, and $T$ the absolute temperature. Spatial discretization of Eq.~(\ref{smoluchowski})~\cite{Bicout1998} results in a master equation that describes the single-particle dynamics between neighbouring intervals along $z$.  Local free energies $F(z)$ and local diffusivities $D_{\perp} (z)$ are then determined self-consistently from the dynamics observed in molecular simulations through Bayesian inference~\cite{Gerhard2005}. 

We apply this formalism to test whether diffusion provides a quantitative description of the single-particle dynamics perpendicular to the confining planes for a fluid confined in a 2d slit pore.  We will also explore whether empirical relations between the fluid density and the diffusion coefficient identified for diffusion in bulk and parallel to the walls $D_\parallel$~\cite{Mittal2006} transfer trivially to perpendicular diffusion. 

Hard spheres (HS) confined between hard walls are arguably the most basic model of confinement.  Nevertheless, essential physics of inhomogeneous fluids is captured, such as pronounced local density variations~\cite{Ted1996} or shifted fluid-solid phase boundaries with respect to bulk~\cite{Marjolein2006}. Also, the unambiguous definition of quantities like accessible volume, and rigorous ways to calculate it for hard sphere fluids~\cite{Widom1963}, can be helpful in elucidating the underlying microscopic mechanisms. Moreover, theoretical and computational predictions for this model system can be tested in experiments on colloids~\cite{Eric2000}. 

We use discontinuous molecular dynamics (DMD) simulations to generate dynamic trajectories for our model system. To simplify the notation, dimensionless quantities will be reported, obtained by appropriate
combinations of a characteristic length (HS particle diameter $\sigma$) and 
time scale ($\sigma \sqrt{m \beta}$, where $m$ is the particle mass). The packing fraction $\phi=\pi\rho / 6$ is defined in terms of the density $\rho$ based on the total (rather than center-accessible) volume~\cite{Mittal2006}. The DMD simulations each involved $N=3000$ identical HS particles.  Periodic boundary conditions were applied in all directions for the bulk fluid and in the $x$ and $y$ directions for the confined fluid. In the confined system, perfectly reflecting, smooth hard walls were placed at $z=\pm H/2$.  
The transverse self-diffusivity $D_{||}$ was obtained by fitting the long-time ($t \gg 1$) behavior of the average mean-squared displacement of the particles to the Einstein relation $\left<\Delta {\bf r}^2\right> = 4D_{||}t$, where $\Delta {\bf r}^2$ corresponds to the mean-square displacement per particle
in the $x$ and $y$ directions. To calculate $D_{\perp}(z)$ for a spatially discretized 
Eq.~(\ref{smoluchowski})~\cite{Gerhard2005}, we use a bin size of 1/10 to divide the space in the $z$ direction. 

\begin{figure}
\scalebox{0.95}{\includegraphics{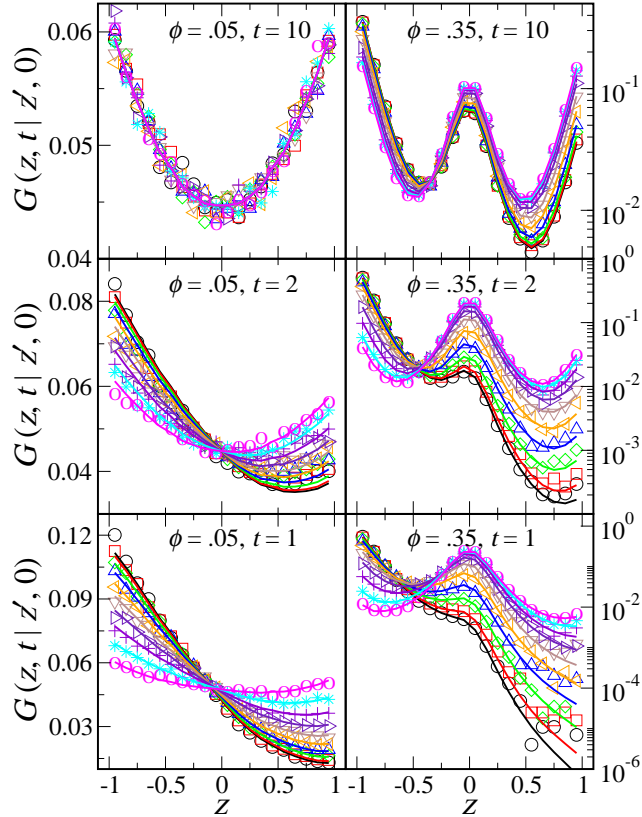}}
\caption{\label{smoul}{The conditional probability $G(z,t|z^{\prime},0)$ of observing a particle at a position $z$ at time $t$ if it started at $z^\prime$ at $t=0$. Results are shown for pore size $H=3$, packing fractions $\phi=0.05$ (left) and 0.35 (right), and times $t=1$, 2, and 10 (bottom to top). The observation time $t=1$ is used to obtain parameters of the diffusion model 
Eq.~(\ref{smoluchowski}) (lines). Simulation results for $G(z,t|z^{\prime},0)$ for different $z^{\prime}$ are shown as symbols, where $z^{\prime}$ varies from $-0.05$ to $-0.95$ (symbol o to 0) in intervals of 0.1. (For reference, mean collision frequencies for the bulk hard-sphere fluid are approximately 852 and 7369 for $\phi$ = 0.05 and 0.35, respectively.)}}
\end{figure}

To test if diffusion captures the dynamics normal to the walls, we compare the long time propagators $G(z,t|z^\prime,0)$ of the Markovian model to the simulation data. $G(z,t|z^\prime,0)$ is the conditional probability that a particle starting from position $z^\prime$ at time 0 is found at $z$ at a later time $t$. As shown in Fig.~\ref{smoul} for two different packing fractions and for a pore size $H=3$, excellent agreement is found between the diffusive model and the MD data over six orders of magnitude in the propagator. 

\begin{figure}
\scalebox{1.}{\includegraphics{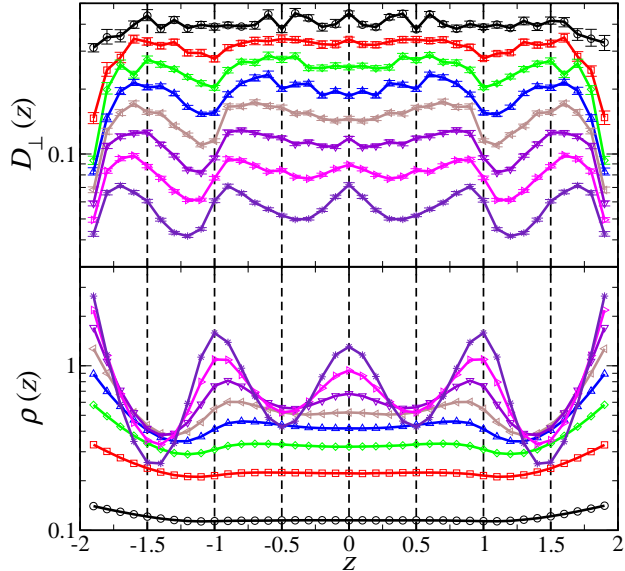}}
\caption{\label{H5-Dz_Rz}{
Local normal diffusivity $D_\perp(z)$ and density profile $\rho(z)$ versus position in the $z$ direction for a pore size $H=5$ and packing fractions $\phi=0.05,$ 0.10, 0.15, 0.20, 0.25, 0.30, 0.35, and 0.40 (top to bottom for $D_\perp(z)$ and bottom to top for $\rho(z)$).}} 
\end{figure}

Figure~\ref{H5-Dz_Rz} compares $D_\perp (z)$ and the local density $\rho(z)$ to explore the effect of wall confinement on the local normal diffusivity. Remarkably, we find that the $D_\perp (z)$ is large where $\rho(z)$ is high (except near the walls, where $\rho(z)$ drops sharply). This low diffusivity near the walls is due to the presence of impenetrable reflective wall boundaries, limiting diffusion to one direction.

At first sight, the positive correlation between high local density
and faster local diffusion may appear counterintuitive as diffusivity for
bulk fluids usually decreases with increasing density. To understand
this unexpected behavior for inhomogeneous fluids, we turn our
attention to the physics of layer formation in the confined
environments and to why the fluid is structured normal to the walls in
the first place. As it turns out, the confined fluid tends to maximize
its entropy by forming these layers~\cite{Kjellander}. Similarly,
the homogeneous hard-sphere fluid crystallizes at high enough
densities to maximize entropy~\cite{Alder1962,Hoover1968}. We note that the
activity $\xi = \exp( \beta \mu) / \lambda^3$ of an equilibrated
confined system is spatially invariant, even though the density is not
(where $\mu$ is the chemical potential and $\lambda$ is the thermal
wavelength).  Also, for a hard-sphere fluid, $\xi = \rho(z)$ /
$P_0(z)$~\cite{Widom1963}, which means that the local insertion
probability $P_0(z)$ (or local available volume~\cite{Sastry1998}) is
directly proportional to the density $\rho(z)$. In other words, the
counterintuitive idea that dense ``layers'' actually have more
available space than the gaps between them is a consequence of this
simple and exact relationship derived more than 40 years ago by Widom~\cite{Widom1963}. 
If we think of diffusion as particles probing their
surroundings for space, then $P_0$ should indeed be a relevant
quantity.  A ``prediction" would then be that $D_\perp(z)$ should
approximately collapse as a function of $P_0(z)$.

\begin{figure}
\scalebox{0.95}{\includegraphics{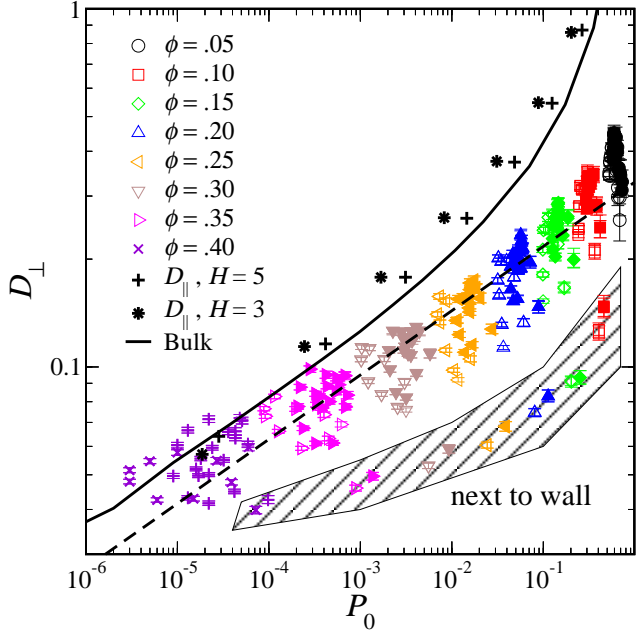}}
\caption{\label{Dz-P0}{Local diffusivity $D_\perp(z)$ at different packing fractions $\phi$ as a function of the local available volume, measured by the local insertion probability $P_0 (z)$. Filled and empty symbols correspond to $H=5$ and 3, respectively. Data for bulk are shown by a solid line. The dashed line is a power law fit to the $D_\perp$ data. The average transverse diffusivity $D_{||}$ is also shown for reference.}}
\end{figure}

To test this prediction, we either need the particle insertion probability $P_0$ as a function of $z$ or the activity $\xi$ as a function of pore size $H$ and average pore density along with the density profile $\rho(z)$. Using the particle insertion method can be very time consuming and even practically limited at high densities. Instead, we use grand canonical transition matrix Monte Carlo (GC-TMMC) simulations to evaluate the functional relationship between the activity $\xi$ and the average pore density $\rho$ for a given $H$. The details of the GC-TMMC method are documented in ~\cite{Errington2003} and the specific implementation details are as in ~\cite{Mittal2007}.

In Fig.~\ref{Dz-P0}, we show $D_\perp$ versus $P_0$ over a wide density range ($\rho=0.05$ - 0.4) for two pore sizes ($H=3$ and 5). The relationship for the bulk hard-sphere fluid and for the transverse diffusivity are also shown on the same plot. We find that the $D_\perp$ data (filled and empty symbols) approximately collapse to a power law form which in turn is very similar to the bulk relationship shown by the solid line. Note that right next to the wall, this relation does not hold (marked by the shaded area in Fig.~\ref{Dz-P0}). However, this is expected because at the wall particles can only diffuse in one direction (i.e., away from the wall). Overall, the approximate collapse in Fig.~\ref{Dz-P0} supports the idea that the local available volume, probed by the insertion probability, is indeed a relevant quantity for diffusion. 

We also notice in Fig.~\ref{Dz-P0} that the local normal diffusivity is lower than the corresponding bulk value, when compared at the same value of $P_0$. This is expected because the presence of the walls directly hinders normal diffusivity in ways that are not reflected by the local available space. Of course, this becomes a relatively small effect at high $\rho$ and is reflected in Fig.~\ref{Dz-P0} by a convergence of the $D_{\perp}$, $D_{||}$, and bulk diffusivity at low $P_0$. The reason for the slightly higher $D_{||}$ stems from a higher chemical potential required to achieve the average pore density equal to the bulk and has been discussed in detail elsewhere~\cite{Mittal2007}.  

\begin{figure}
\scalebox{1.0}{\includegraphics{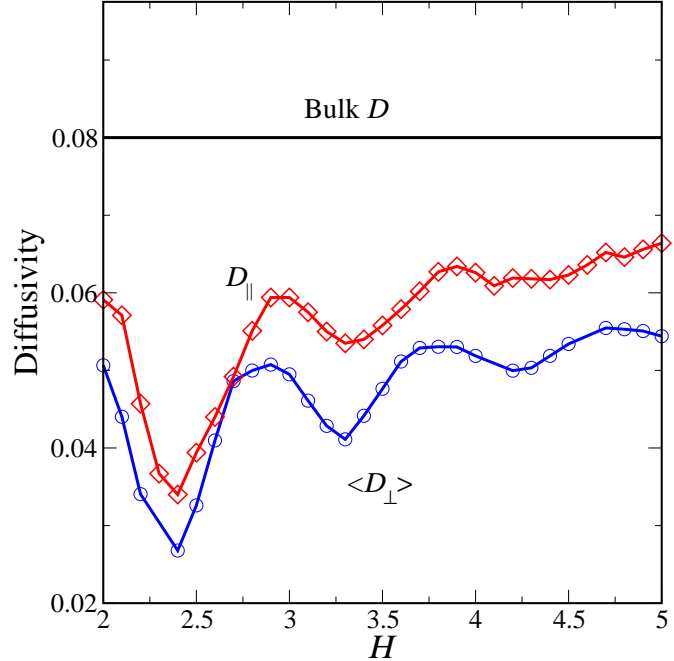}}
\caption{\label{Dz_Dt}{The average normal (blue circles) and transverse (red diamonds) diffusivity versus the pore size $H$ is shown for a packing fraction of $\phi=0.40$. The corresponding bulk diffusivity is shown as a solid horizontal line.}}
\end{figure}

We can also use our formalism to explore the coupling between 
diffusion in the transverse and normal directions. 
In an earlier study~\cite{Mittal2007}, it was observed that 
$D_{||}$ of the confined HS fluid shows pronounced
negative deviations from bulk fluid behavior for relatively high density 
(e.g., $\phi = 0.4$) and small pore widths ($H<3$).  These
deviations have an oscillatory dependence on $H$ with slower diffusion
occuring for pore sizes that do not naturally accomodate an
integer number of particle ``layers'' in the density profile.   
Based on such information, it was
hypothesized~\cite{Mittal2007} that one might also expect a coupling between 
single-particle dynamics in directions parallel and normal to the
confining walls.  Here, we test this idea
by calculating the average normal diffusivity 
$\langle D_\perp\rangle = \int_0^{H/2} D_\perp \rho {\text {dz}} / \int_0^{H/2} \rho {\text
  {dz}}$ and comparing it to $D_{||}$ in
Fig.~\ref{Dz_Dt}. As can be seen, the $H$-dependent oscillation of
$\langle D_\perp \rangle$ closely tracks that of $D_{||}$.  
Remarkably, the oscillations in $D_{||}$ and $\langle D_\perp\rangle$ as a function of $H$ 
follow the pore-width dependence of the phase boundary $\phi_{\mathrm f}(H)$ between the 
confined solid and fluid HS system~\cite{Marjolein2006}, and also the excess entropy per 
particle~\cite{Mittal2007}. These correlations illustrate the connection between dynamic 
and thermodynamic properties arising from packing frustration.  
In particular, these results provide strong
support for the picture that confinement-induced frustration 
reduces diffusive particle motions in both principal directions.

\begin{acknowledgments}
J.M. wishes to thank Jyoti Seth (University of Texas at Austin) for her careful reading of an early draft. This research was supported in part by the Intramural Research Program of the NIH, NIDDK. T.M.T. and J.R.E. acknowledge the financial support of the National Science Foundation under Grant Nos. CTS-0448721 and CTS-028772, respectively. T.M.T. also acknowledges the support of the David and Lucile Packard Foundation and the Alfred P. Sloan Foundation. The Texas Advanced Computing Center (TACC) and
University at Buffalo Center for Computational Research provided computational resources for this study. A portion of this study utilized the high-performance computational capabilities of the Biowulf PC / Linux cluster at the National Institutes of Health, Bethesda, MD (http://biowulf.nih.gov).
\end{acknowledgments}

\end{document}